\documentclass[useAMS,usenatbib]{mn2e}
\usepackage{epsfig,amsmath,amssymb,amsfonts,mathrsfs,latexsym,graphicx}

\usepackage[draft]{hyperref}
\usepackage{natbib} 
\usepackage{color}
\usepackage{algorithm}
\usepackage{ textcomp } 

\usepackage{fixltx2e}
\usepackage{amsmath}
\usepackage{float}
\usepackage[caption = false]{subfig}
\usepackage{multirow}
\include{journals}

\graphicspath{{images/}}

\newcommand{\Nif}{$^{56}$Ni\,}

\newcommand{\xone}{x_1}

\newcommand{\covariates}{X}
\newcommand{\regcoeff}{\mathscr{B}}
\newcommand{\zhat}{\hat{z}}
\newcommand{\mB}[1]{{m}^\star_{B#1}}
\newcommand{\Cparams}{\mathscr{C}} 
\newcommand{\unif}{{\hbox{\sc Uniform}}}

\newcommand{\Mnot}{M_0}
\newcommand{\normal}{\mathcal{N}}
\newcommand{\invgamma}{{\hbox{\sc InvGamma}}}
\newcommand{\sigmares}{\sigma_{\rm res}}
\newcommand{\sigmapec}{\sigma_{\rm pec}}
\newcommand{\xstar}{x_{1\star}}
\newcommand{\Rx}{R_{x_1}}
\newcommand{\tstar}{t_{2\star}}
\newcommand{\Rt}{R_{t_2}}
\newcommand{\cstar}{c_\star}
\newcommand{\Rc}{R_c}
\newcommand{\logunif}{{\hbox{\sc LogUniform}}}
\newcommand{\tslope}{\gamma}
\newcommand{\lm}{$L_{\rm max}$}

\newcommand{\apj}{ApJ}
\newcommand{\apjl}{ApJL}

\newcommand{\aj}{AJ}
\newcommand{\aap}{A\&A}
\newcommand{\mnras}{MNRAS}

\newcommand{\pasp}{PASP}

\def\gsim{\mathrel{\rlap{\lower 4pt \hbox{\hskip 1pt $\sim$}}\raise 1pt \hbox {$>$}}}
\def\lsim{\mathrel{\rlap{\lower 4pt \hbox{\hskip 1pt $\sim$}}\raise 1pt \hbox {$<$}}}
\newcommand{\BAHAMAS}{{\em BAHAMAS}}

\title[SNIa NIR standardization]{Standardizing Type Ia supernovae using Near Infrared rebrightening time}

\author[]
{\parbox{\textwidth}{\vspace{-.5cm} \large  
H.~Shariff$^{1,2}$\thanks{E-mail: 
hikmatali.shariff11@imperial.ac.uk},
S.~Dhawan$^{3,4,6}$, 
X.~Jiao$^{2,7}$,
B.~Leibundgut$^{3,4}$, 
R.~Trotta$^{1,2,8}$,
D.~A.~van Dyk$^{2,7,8}$
}\vspace{0.6cm}\\
\parbox{\textwidth}{ 
$^{1}$ Astrophysics Group, Physics Department, Imperial College London, Prince Consort Rd, London SW7 2AZ \\ 
$^{2}$ Imperial Centre for Inference and Cosmology, Blackett Laboratory, Prince Consort Rd, London SW7 2AZ\\
$^{3}$ European Southern Observatory, Karl-Schwarzschild-Strasse 2, D-85748 Garching bei M\"unchen, Germany \\
$^{4}$  Excellence Cluster Universe, Technische Universit\"at M\"unchen, Boltzmannstrasse 2, D-85748, Garching, Germany\\
$^{6}$ Physik Department, Technische Universit\"at M\"unchen, James-Franck-Strasse 1, D-85748 Garching bei M\"unchen\\
$^{7}$ Statistics Section, Mathematics Department,  Huxley Building, Imperial College London,  London SW7 2AZ\\
$^{8}$ Data Science Institute, William Penney Laboratory, Imperial College London, London SW7 2AZ\\
}}

\begin{document}

\date{Accepted ... Received ...; in original form ...}

\pagerange{\pageref{firstpage}--\pageref{lastpage}} \pubyear{2015}

\maketitle

\begin{abstract}

Accurate standardization of Type Ia supernovae (SNIa) is instrumental to the usage of SNIa as distance indicators. We analyse a homogeneous sample of 22 low-$z$ SNIa, observed by the Carnegie Supernova Project (CSP) in the optical and near infra-red (NIR). We study the time of the second peak in the NIR band due to rebrightening, $t_2$, as an alternative standardization parameter of SNIa peak brightness. We use \BAHAMAS, a Bayesian hierarchical model for SNIa cosmology, to determine the residual scatter in the Hubble diagram.

We find that in the absence of a colour correction, $t_2$ is a better standardization parameter compared to stretch: $t_2$ has a $1 \sigma$ posterior interval for the Hubble residual scatter of $\sigma_{\Delta \mu} = \{0.250,0.257 \}$ , compared to $\sigma_{\Delta \mu} = \{0.280,0.287 \}$ when stretch ($x_1$) alone is used. We demonstrate that when employed together with a colour correction, $t_2$ and stretch lead to similar residual scatter.
Using colour, stretch and $t_2$ jointly as standardization parameters does not result in any further reduction in scatter, suggesting that $t_2$ carries redundant information with respect to stretch and colour. With a much larger SNIa NIR sample at higher redshift in the future, $t_2$ could be a useful quantity to perform robustness checks of the standardization procedure. 

\end{abstract}


\section{Introduction}
Type Ia supernovae (SNIa) are exceptionally useful distance indicators in cosmology and have been instrumental in the discovery of the accelerated expansion of the universe \citep{1998AJ....116.1009R, 1999ApJ...517..565P}. Their apparent magnitudes must be ``standardized'' by correcting for empirical correlations between observable properties and intrinsic magnitude in order to reduce residual scatter sufficiently to measure cosmological parameters \citep{1993ApJ...413L.105P, 1996ApJ...473..588R, 1998A&A...331..815T}. 

To date, cosmological parameter constraints from SNIa have been derived from optical lightcurve data. There are however clear indications that near infra-red (NIR) data improve precision and accuracy further~\citep{2011ApJ...731..120M}. At NIR wavelengths (900 nm $< \lambda <$ 2000 nm) SNIa exhibit more uniform brightness, without the need for empirical standardization \citep{2004ApJ...602L..81K,2008ApJ...689..377W,2009ApJ...704..629M, 2011ApJ...731..120M}. 
The scatter in the
peak luminosity in these studies can be as low as 0.1 mag. Furthermore, NIR light is less affected by extinction due to dust, which makes NIR data less prone to pernicious dust-related systematics. Distances in the NIR can be measured to better than 6$\%$ precision, making this wavelength region an exciting prospect for SNIa cosmology \citep{2012PASP..124..114K}. 

The lightcurve morphology in the NIR is markedly different from that in the optical, showing a pronounced second
maximum in $IYJHK$ filters for ``normal'' SNIa \citep{1981ApJ...251L..13E,1988PhDT.......171L, 1996AJ....112.2438H,2010AJ....139..120F}. This rebrightening offers interesting clues into the physics of the explosion. The timing of the second maximum ($t_2$, measured as the time between $B$-band maximum light and the second maximum in a given NIR filter) in the $YJHK$ filters is strongly correlated with the optical decline rate of SNIa (measured by the $\Delta m_{15}$ parameter) as shown in \citet{2012A&A...537A..57B} and \citet{2015MNRAS.448.1345D}. \citet{2006ApJ...649..939K}, based on radiative transfer calculations of synthetic lightcurves, predicted that $t_2$ is a function of the \Nif mass produced by the SNIa. Indeed, \citet{2015suhail} found a strong correlation between the peak bolometric luminosity \citep[and therefore, \Nif mass, see][]{1982ApJ...253..785A} and $t_2$ for a sample of SNIa with low-extinction from host galaxy dust. They demonstrate that this parameter can be used to infer the \Nif mass for SNIa. 


There are many reasons why $t_2$ is a potentially useful quantity for SNIa cosmology.  Firstly, since $t_2$ is a timing (and not a flux) estimate, it is unaffected by host galaxy reddening, which is not the case for optical decline rate parameters, e.g. $\xone$ and $\Delta m_{15}$ \citep[see][for a discussion]{2011ApJ...735...20A}. Secondly, the rebrightening in the NIR is due to an ionization transition of Fe-group elements from doubly to singly ionized atoms at a transition temperature $\sim$ 7000 K~\citep[see][]{2006ApJ...649..939K,2015MNRAS.448.2766B}. The time at which this transition occurs is driven by the amount of heating from \Nif\ produced by the explosion. Since  \Nif\ is the primary energy source for the peak brightness of the SNIa~\citep{1982ApJ...253..785A}, $t_2$ would be expected to correlate strongly with the peak magnitude of the SNIa and hence be an effective parameter for standardization. 

Thirdly, the rebrightening in the NIR is an exclusive feature of SNIa, not observed in any other type of SN. This makes the second maximum a useful indicator to distinguish SNIa from other types of SNe in absence of spectroscopic confirmation. Therefore, observing the NIR second maximum could become critical for photometric classification of SNIa in future SN surveys. Given that such data should become available with future facilities --e.g. Euclid (expected launch 2019), JWST (expected 2018), WFIRST \citep[expected launch $\sim$ 2020; see][for a review]{2013RSPTA.37120282H}-- studying the standardization properties of $t_2$ can provide an alternative and more robust route to SNIa cosmology at no additional observational cost. Furthermore, the time-delay between the optical peak brightness and NIR rebrightening means that NIR observations can be scheduled without the need for prompt alerts required to sample densely the optical lightcurve near its peak, which is necessary for an accurate estimate of $\Delta m_{15}$.

The aim of this work is to compare the residual scatter in magnitudes after empirical standardization for a sample of nearby SNIa using the traditional stretch parameter ($\xone$) versus what can be obtained when using $t_2$ instead (or in addition). We address the question of whether the NIR rebrightening time can be used to reduce the residual scatter in the peak luminosity of SNIa. This paper is structured as follows. In section~\ref{sec-meth} we describe the data set we use and present the Bayesian methodology for our analysis including a description of the parameters of interest. We present the results of applying our method to this low-$z$ SNIa data set in section~\ref{sec-res} and conclude in section~\ref{sec-conc}.

\section{Methodology}
\label{sec-meth}
\subsection{Data}
We have compiled a sample of 22 SNIa, in the redshift range $0.01 \leq z \leq 0.047$, all with well-sampled optical and NIR photometry. The source of NIR SNIa photometry is the Carnegie Supernova Project \citep[CSP;][]{2010AJ....139..519C,2011AJ....142..156S}. The low-$z$ CSP provides a sample of SNIa with optical and NIR light curves in a homogeneous and well-defined photometric system (the Vega magnitude system) and thus forms an ideal base for the evaluation of light curve properties. 

The SALT2 fit parameters, i.e. peak B-band magnitude $m_B$, lightcurve stretch correction $\xone$, and colour correction $c$, ~\citep{2007A&A...466...11G} are taken from the analysis in~\citet{2014ApJ...795...44R}. The total number of SNIa with NIR $t_2$ estimates is larger than presented here, however, we only use the subset analysed as part of the low-$z$ anchor in \citet{2014ApJ...795...44R}. The NIR rebrightening time, $t_2$, in the $J$-band, is evaluated as described in \citet{2015MNRAS.448.1345D} and subsequently re-centered to the sample mean value, $\langle t_2 \rangle = 27.96$ days.

\subsection{Setup and Method}

In order to determine the standardization parameters and residual dispersion in the SNIa magnitudes, we use the Bayesian hierarchical method \BAHAMAS\ \citep{2015arXiv151005954S}. 
\citet{2011MNRAS.418.2308M} introduced Bayesian hierarchical modeling to the problem of cosmological parameter extraction from SALT2 fits. 
The key feature is the hierarchical treatment of sources of uncertainty, comprising of both measurement errors and population variability. Each observed covariate (for example $\hat{m}_{Bi}$, the observed apparent magnitude of the $i$-th SNIa) is assumed to have an underlying true (latent) value (for example $m_{Bi}$, the real apparent magnitude of the $i$-th SNIa) that is unobserved. Linear regression is then applied to the true value of the covariate, which itself is drawn probabilistically from a distribution describing the population of SNIa.  Finally, latent values are marginalized from the posterior distribution. \citet{2015arXiv151005954S} further developed this Bayesian approach to include additional (or alternative) covariates for the standardization of SNIa, and to provide explicit sampling of the latent variables. We refer to \citet{2015arXiv151005954S} for full details about the hierarchical model and the sampling methods (see also \cite{Nielsen:2015pga} for a similar model but applied under a frequentist framework; \cite{Rubin:2015rza} for a different implementation of a similar hierarchical Bayesian model, and \cite{Ma:2016sio} for an analysis using Bayesian graphs). 

Another feature that distinguishes this Bayesian procedure from the standard $\chi^2$ approach is the treatment of the absolute magnitude of SNIa's. Rather than assuming that all SNIa's have the same intrinsic magnitude (and then inflate the observational errors to obtain $\chi^2/$dof =  1, as in the standard approach), each SNIa is assigned its own absolute magnitude, $M_i^\epsilon$. These are assumed to follow an underlying Gaussian distribution (denoted by $\normal$), $M_i^\epsilon \sim \normal (M_0, \sigmares^2)$, whose mean, $M_0$, and residual dispersion, $\sigmares$, are determined from the data. 
Similarly, the other latent parameters are modeled hierarchically as
$x_{1i} \sim \normal (\xstar,R_{x_1})$, $t_{2i},
\sim \normal (\tstar,R_{t_2})$
and $c_i \sim \normal (\cstar,R_c)$, i.e., the aforementioned distributions describing the population of SNIa, where the means and standard deviations of the distributions are Bayesianly determined from the data.

This hierarchical structure has an advantage of ``borrowing strength'', in particular, when estimating the intrinsic (and unaccounted for) scatter of SNIa ($\sigmares$).
\citet{2011MNRAS.418.2308M} also showed that this method reduces the mean squared error of the parameter estimators when compared to the standard approach.
In this work, we extend \BAHAMAS\ to include  $t_2$ as an additional linear covariate (or as an alternative to $x_1$).

The generalized Phillips corrections, including any number of (linear) covariates can be written as: 
\begin{equation} 
\label{eq:covariates_relation_generalized}
\mB{i} =\mu_{i}(\zhat_i, \Cparams) + \covariates_i^T \regcoeff +M_i^{\epsilon},
\end{equation}
where $\mu_{i}(\zhat_i, \Cparams)$ is the distance modulus at the observed redshift $\zhat_i$ for cosmological parameters $\Cparams$, $\covariates_i$ is a vector of covariates, and $\regcoeff$ is the vector of regression coefficients. In the standard SALT2 analysis $\covariates_i = \{x_{1i},c_i\}$ and $\regcoeff =\{-\alpha,\beta\}$, where $\alpha$ is the slope of the stretch correction parameter and $\beta$ is the slope of the colour correction parameter. (Everywhere, hats denote measured quantities.) In this analysis we consider 5 cases:  
\begin{itemize}
  \item $\covariates_i = \{x_{1i}\}$ and $\regcoeff =\{-\alpha\}$
  \item $\covariates_i = \{t_{2i}\}$ and $\regcoeff =\{-\tslope\}$
  \item $\covariates_i = \{x_{1i},c_i\}$ and $\regcoeff =\{-\alpha,\beta\}$
  \item $\covariates_i = \{t_{2i},c_i\}$ and $\regcoeff =\{-\tslope,\beta\}$
  \item $\covariates_i = \{x_{1i},t_{2i},c_i\}$ and $\regcoeff =\{-\alpha,-\tslope,\beta\}$,
\end{itemize}
where $\tslope$ is the NIR rebrightening parameter, giving the (negative of the) slope of the linear relationship between $t_2$ and intrinsic magnitude. 

\begin{table*}\footnotesize
\begin{center}
\begin{tabular}{ll} 
\hline\hline 
Parameter  & Notation and Prior Distribution\\ 
\hline 
\multicolumn{2}{c}{Covariates} \\\hline
Negative of the coefficient of stretch covariate  & $\alpha \sim \unif(-1,1)$ \\ 
Coefficient of colour covariate & $\beta  \sim \unif(-4,4)$ \\ 
Negative of the coefficient of NIR rebrightening time  & $\tslope  \sim \unif(-1,1)$ \\ 
\hline 
\multicolumn{2}{c}{Population-level distributions} \\\hline
Mean of absolute magnitudes  & $\Mnot^{\epsilon} \sim \normal(-19.3,2^2)$ \\ 
Residual scatter after corrections  & $\sigmares^2 \sim \invgamma(0.003,0.003)$ \\ 
Mean of stretch & $\xstar \sim \normal(0,10^2)$\\ 
SD of stretch & $\Rx \sim \logunif(-5,2)$ \\ 
Mean of $t_2$  & $\tstar \sim \normal(0,10^2)$\\ 
SD of $t_2$  & $\Rt \sim \logunif(-5,2)$ \\ 
Mean of colour  & $\cstar \sim \normal(0,1^2)$  \\ 
SD of colour  & $\Rc \sim \logunif(-5,2)$ \\ 
\hline \hline
\end{tabular}
\caption{Summary of the parameters, notations, and prior distributions used in our hierarchical model.  ``SD'' stands for ``standard deviation''.  See \citet{2015arXiv151005954S} for more details.
\label{table:main_params}
}
\end{center}
\end{table*}%

We adopt a flat $\Lambda$CDM cosmology with fixed cosmological parameters, $ \Cparams  = \{  \Omega_m = 0.3, w = -1, H_0 = 70 \}$, where $\Omega_m$ is the matter density parameter, $w$ the dark energy equation of state parameter and $H_0$ the Hubble parameter today. We fix the cosmology since this parameters are unconstrained by low redshift SNIa data alone.

The priors for the other parameters in our model are given in Table \ref{table:main_params}. A significant fraction of the residual dispersion in low-$z$ SNIa is due to peculiar velocities, since these objects are not fully in the Hubble flow. To account for the variance due to peculiar velocities, we follow a procedure similar to \citet{2009ApJ...704..629M}. Specifically, we add a term, $\sigma_{\mu}^2$, to the apparent magnitude error; $\sigma_\mu^2$ depends on the peculiar velocities uncertainty, $\sigmapec$, and redshift measurement error, $\sigma_z$, as
\begin{equation}
\label{eq:peculiar_corrections}
\sigma_{\mu,i}^2 = \left( \frac{5}{\hat{z}_i \ln(10)}  \right )^2 \left[ \sigma_{z,i}^2+\frac{\sigmapec^2}{c^2} \right].
\end{equation} 
Following \cite{2009ApJ...704..629M}, we set $\sigma_\text{pec}$ = $150$km/s.

To ensure the robustness of the posterior distribution, we vary the choice of prior distribution for the residual intrinsic dispersion, $\sigma_\text{res}$, using \invgamma(0.1,0.1), \invgamma(0.03,0.03), \invgamma(0.003,0.003) and  $\logunif(-5,2)$; $\invgamma(a,b)$ denotes a random variable whose reciprocal follows a Gamma distribution\footnote{More specifically: if $X \sim \invgamma (a,b)$, its probability density is given by $p(x)=\frac{b^a}{\Gamma(a)}x^{-a-1}e^{-b/x}$.} with mode equal to $\frac{b}{a+1}$. 
While the posteriors for all other parameters are fairly  independent of the choice of prior distribution for $\sigma_\text{res}$, 
the posterior distribution of $\sigma_\text{res}$ is sensitive to this choice. (This is not unexpected given the small number of SNIa in our sample.) Thus we do not use the posterior distribution of $\sigma_\text{res}$ to quantify and compare the residual scatter for different models. Instead, we quantify the residual scatter in the Hubble diagram with the Hubble residual between the observed distance modulus, $\hat{\mu}_i(M_0,\regcoeff) =   \hat{m}_{Bi} - \hat{{\covariates_i}}^T \regcoeff - M_0 $, and the model distance modulus, $\mu_(\zhat,\Cparams)$, for SNIa $i$, that is,  
\begin{equation}
\label{eq:hubble_residuals}
\Delta \mu_i = \hat{\mu}_i(M_0,\regcoeff) - \mu(\zhat_i,\Cparams).
\end{equation} 

We emphasize that $\hat{\mu}_i(M_0,\regcoeff)$ is a function of the unknown parameters $M_0$ and $\regcoeff$. Thus in a Bayesian analysis $\hat{\mu}_i(M_0,\regcoeff)$ itself has a posterior distribution. 
We then calculate the posterior distribution of the sample standard deviation of the $\Delta \mu_i$, i.e. 
\begin{equation}
\label{eq:s_deltamu}
\sigma_{\Delta \mu} = \sqrt{\frac{1}{n-1}\sum_{i=1}^n(\Delta \mu_i - \overline{\Delta\mu})^2 },
\end{equation} 
where $\overline{\Delta \mu} = \frac{1}{n}\sum_{i=1}^n \Delta \mu_i$. 
Because $\hat{\mu}_i(M_0,\regcoeff)$ is a function of the unknown parameters $M_0$ and $\regcoeff$, $\sigma_{\Delta \mu}$ is also a function of $M_0$ and $\regcoeff$ and itself has a posterior distribution. Thus we calculate the posterior distribution of $\sigma_{\Delta \mu}$.
We find that $\sigma_{\Delta \mu}$ is independent of the prior choice (of $\sigma_\text{res}$) and quantifies the residual scatter in the Hubble diagram well. We use $\sigma_{\Delta \mu}$ to compare the performance of the five choices of covariates enumerated in the five cases mentioned previously.

In order to cross-check our numerical results, we obtain samples from the joint posterior distribution of the parameters of interest using both a Gibbs sampler and a Metropolis-Hastings algorithm, obtaining identical results up to Monte Carlo noise. (Details on the sampling algorithms can be found in \citet{2015arXiv151005954S}.)
We marginalize out the latent variables (via Monte Carlo for the Gibbs sampler and analytically for the Metropolis-Hastings algorithm) and present marginal posterior distributions for the parameters of interest, including $\sigma_{\Delta \mu}$.

\section{Results}
\label{sec-res}

Table~\ref{table:posterior}  presents the posterior mean and standard deviation of the regression coefficients, as well as of $\Mnot ^{\epsilon}$ for each of the five cases. 
When $t_2$ is used as the sole covariate, the (negative of the) slope of the linear relationship with intrinsic magnitude, $\tslope$, shows a 3.4$\sigma$ deviation from zero. When colour is added as an additional covariate, the significance of the $t_2$ coefficient increases to $>5\sigma$. When all three covariates are used together, the (absolute) values of both the rebrightening time and the stretch correction slopes reduce, suggesting that (as expected) $t_2$ and $x_1$ encode similar standardization information.

\begin{table*}\footnotesize
\begin{center}
\begin{tabular}{llllll} 
\hline\hline 
Parameter  & \multicolumn{5}{c}{Covariates}\\ 
& $x_1$ & $t_2$ & $x_1$ and $c$ & $t_2$ and $c$ & $x_1$, $t_2$ and $c$ \\
\hline 
$\alpha$ &$0.139 \pm 0.060$ & n/a & $0.168 \pm 0.035$ & n/a & $0.039 \pm 0.069$   \\ 
$\tslope$ & n/a & $0.037 \pm 0.011$ & n/a & $0.037 \pm 0.007$ & $0.029 \pm 0.014$ \\ 
$\beta$ & n/a & n/a & $3.238  \pm 0.560$ & $2.819  \pm 0.493$ & $2.948  \pm 0.534$  \\ 
\hline 
$\Mnot^{\epsilon}$ & $-19.428\pm 0.070$ & $-19.345 \pm 0.056$ & $-19.313 \pm 0.045$ & $-19.231 \pm 0.040$ & $-19.250 \pm 0.053$\\ 
\hline \hline
\end{tabular}
\caption{Posterior means and 1$\sigma$ marginal posterior intervals for the regression coefficients and population parameters for the five models considered.}
\label{table:posterior}
\end{center}
\end{table*}%

\begin{table*}\footnotesize
\begin{center}
\begin{tabular}{lllllll} 
\hline\hline 
Covariates  & \multicolumn{5}{c}{Percentile}  \\ 
& 5th & 25th & 50th & 75th & 95th  & $1 \sigma$ Interval \\
\hline 
$x_1$ &0.281 & 0.281 & 0.284 & 0.290 & 0.309 & $0.280 - 0.287$  \\ 
$t_2$ &0.250 & 0.251 & 0.254 & 0.260 & 0.279 & $0.250 - 0.257$ \\ 
$x_1$ and $c$ &0.159 & 0.162 & 0.167 & 0.174 & 0.191 & $0.159 - 0.171$  \\ 
$t_2$ and $c$ &0.148 & 0.150 & 0.154 & 0.161 & 0.178 & $0.147 - 0.159$ \\ 
$x_1$,$t_2$ and $c$  &0.145 & 0.150 & 0.155 & 0.164 & 0.184 & $0.144 - 0.161$ \\ 
\hline \hline
\end{tabular}
\caption{Posterior percentiles for the Hubble diagram residual scatter,  $\sigma_{\Delta \mu}$, for the five models considered. Also included is the shortest $1 \sigma$ interval, which extends from the maximum to the 68.3th percentile of the posterior distribution.}
\label{table:sdmu}
\end{center}
\end{table*}%

\begin{figure*}
\includegraphics[width=.8\textwidth]{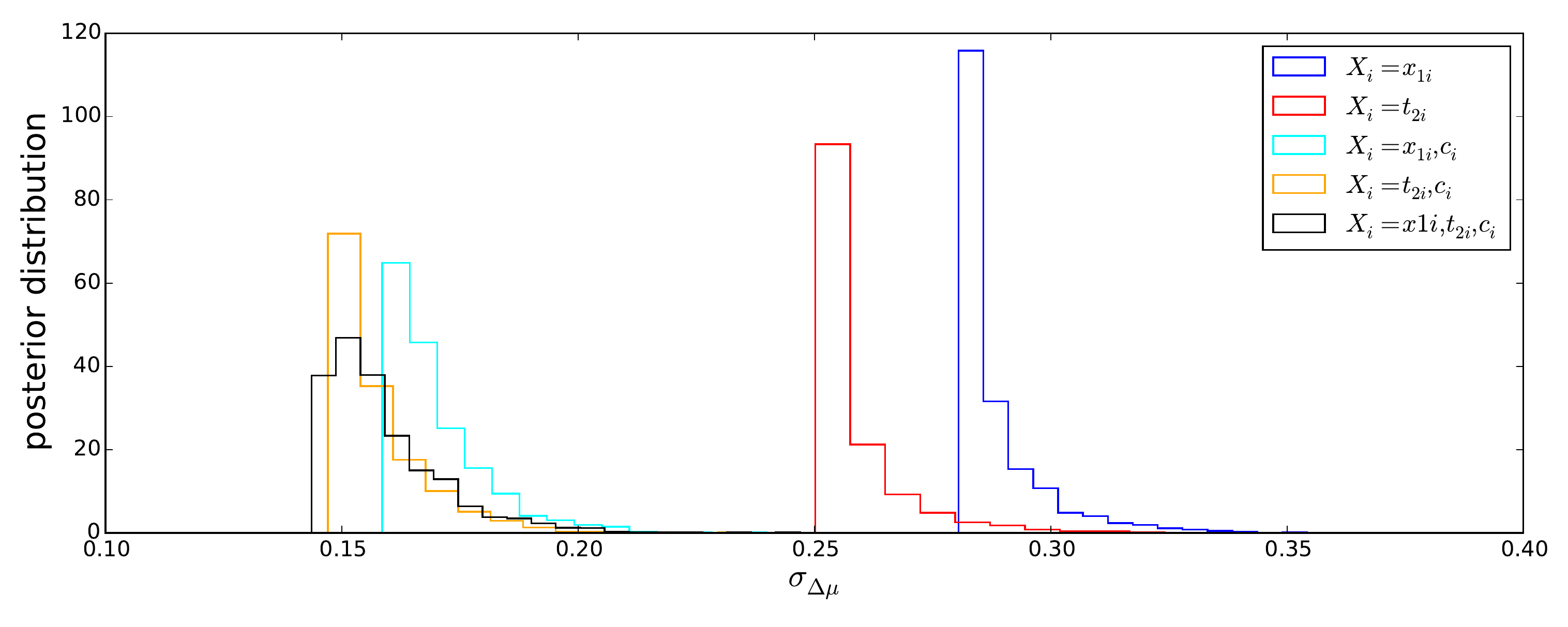}
\caption{Marginal posterior distributions of the Hubble diagram residual scatter, $\sigma_{\Delta \mu}$, for all the five cases considered. Blue colour is for the case using only the stretch correction, $x_1$, as standardization covariate, while red is for using only the NIR rebrightening time, $t_2$. Cyan is for including both stretch and colour, while orange is for using both rebrightening time and colour. Black is for the case when all three covariates are used. Posteriors are normalised.}
\label{fig:s_deltamu}
\end{figure*}

\begin{figure*}
\includegraphics[width=.8\textwidth]{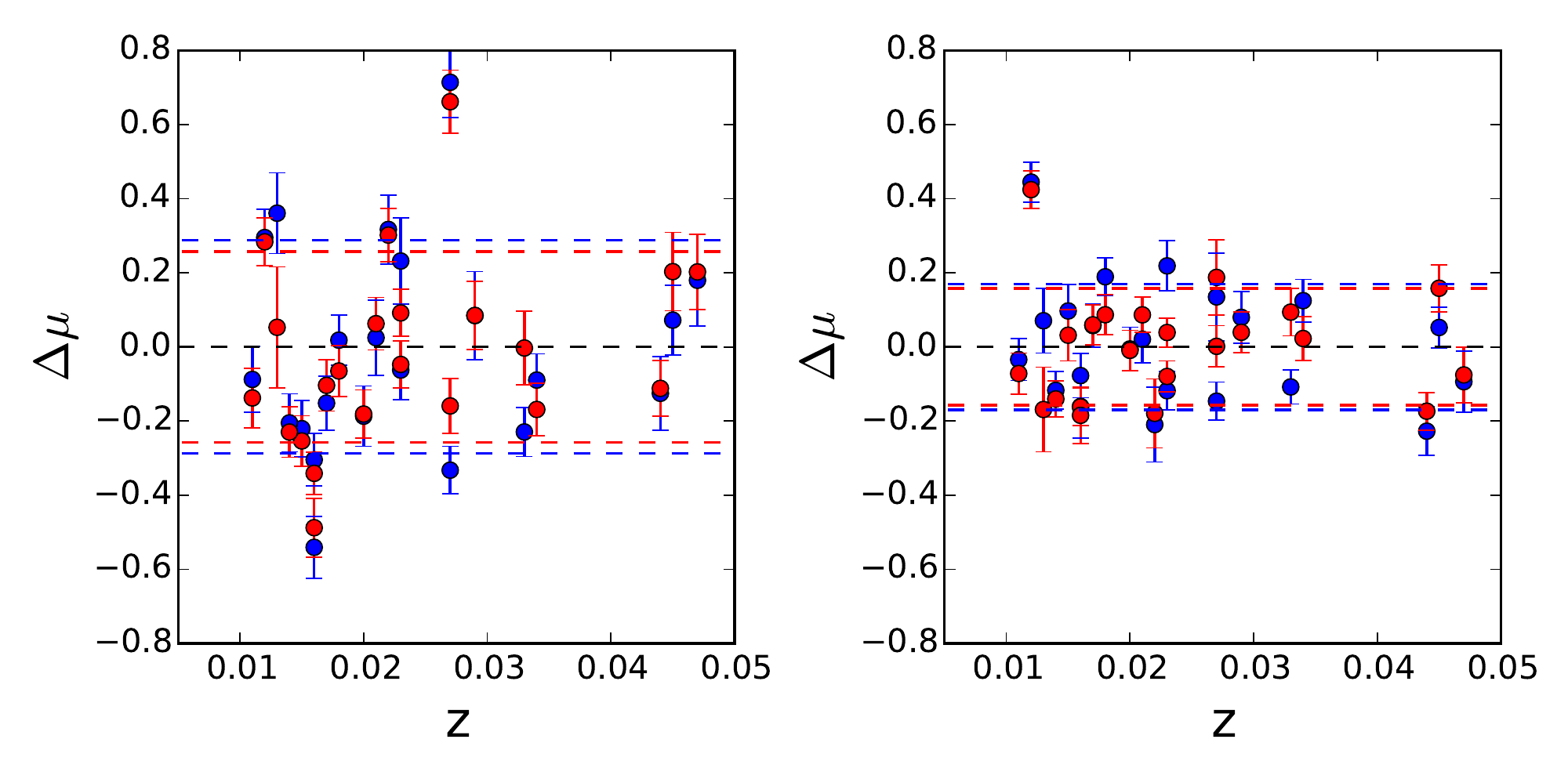}
\caption{Hubble residuals
as a function of redshift. Blue colour is for the cases using the stretch parameter, $x_1$, as a standardization covariate, while red is for using NIR rebrightening time, $t_2$, as a covariate. The left (right) panel excludes (includes) colour correction as a covariate.
Errorbars are the posterior standard deviation of $\Delta \mu$. The dashed red/blue lines indicate the posterior mean of $\sigma_{\Delta \mu}$.}
\label{fig:deltamu}
\end{figure*}

Since the posterior distributions of $\sigma_{\Delta \mu}$ are highly skewed, Table~\ref{table:sdmu} reports their 5th, 25th, 50th, 75th, and 95th percentiles,  rather than their means and the standard deviations. 
Also shown are $1 \sigma$ posterior intervals for each distribution; we choose the shortest $1\sigma$ intervals which in this case extend from the minimum to the 68th percentile of each distribution.
The marginal posterior distributions of $\sigma_{\Delta \mu}$ are displayed in Figure \ref{fig:s_deltamu} for all 5 cases considered. 
Based on the residual scatter in the Hubble diagram, $\sigma_{\Delta \mu}$, in the absence of colour correction, $t_2$ alone is a better standardization quantity than stretch:  the (shortest) $1 \sigma$ posterior interval is $\sigma_{\Delta \mu} = \{0.250,0.257 \}$ for the former, while it is $\sigma_{\Delta \mu} = \{0.280,0.287 \}$ for the latter. This can further be seen in the left panel of Figure~\ref{fig:deltamu}. On average, $t_2$ (red) leads to smaller Hubble residuals than $x_1$ (blue). However, when the colour correction is added to the regression, NIR  rebrightening time and stretch lead to similar residual scatter, with rebrightening time and colour ($1 \sigma$ posterior interval is $\sigma_{\Delta \mu} = \{0.149,0.159 \}$) performing slightly better than stretch and colour ($1 \sigma$ posterior interval is $\sigma_{\Delta \mu} = \{0.159,0.171 \}$), which is also shown in the right panel of Figure~\ref{fig:deltamu}.
When using all three covariates together, we observe no further reduction in the residual scatter. We conclude that $t_2$ can be effectively used as an alternative covariate to $x_1$ for the standardization of SNIa, but does not lead to further reduction in the residuals scatter in the Hubble diagram once both stretch and colour corrections have been included as covariates.
We also consider the case when colour is the only covariate, and find that the posterior distribution of $\sigma_{\Delta \mu}$ is comparable to the case when $t_2$ is the only covariate.

As shown in Figure~\ref{fig:s_deltamu}, the posterior distributions of $\sigma_{\Delta \mu}$ in all five cases are skewed with long tails for larger values and sharp lower bounds.
The sharp lower bound is a feature of the likelihood function, rather than an artifact of the analysis, or a prior-dependent feature. In order to show this, for each of the 5 cases considered, we compute the maximum likelihood value of $\sigma_{\Delta \mu}$ by optimizing $\{ M_0, \regcoeff \}$. We find that in all five cases, the maximum likelihood value of $\sigma_{\Delta \mu}$ coincides with the lower bound of its posterior distribution. With a Gaussian likelihood, the maximum likelihood value of the residual variance is formed by minimizing the sum of squared residuals over the possible values of the regression coefficients. This means that there are {\it no values} of $\{ M_0, \regcoeff \}$ that produce values of $\sigma_{\Delta \mu}$ less than the lower bounds in each of the five posterior distributions. As defined in Equation~\eqref{eq:s_deltamu}, $\sigma_{\Delta \mu}$ is a function of $M_0$ and $\regcoeff$ and its posterior distribution is determined by theirs. 

\section{Conclusions}
\label{sec-conc}

We demonstrated on a low-$z$ SNIa sample that the waiting time for NIR rebrightening, $t_2$, is significantly better at calibrating the peak magnitude of SNIa when compared with stretch alone.
\citet{2015MNRAS.448.1345D} found a correlation between $t_2$, $\Delta m_{15}$ and the time of maximum $(B-V)$ colour (denoted by $t_L$). They inferred that the diversity in $t_2$ values of SNIa is driven by different masses of \Nif produced in the explosion. As a follow-up, \citet{2015suhail} identified a strong correlation between the peak bolometric luminosity (\lm) and $t_2$ for a sample of SNIa with well-measured NIR data. Since the luminosity at peak corresponds to the instantaneous energy deposition rate from \Nif decay \citep[known as ``Arnett's Rule"][]{1982ApJ...253..785A}, the authors used the correlation between \lm\ and $t_2$ and Arnett's rule to infer a \Nif mass distribution. The correlation between \Nif mass and $t_2$ is stronger than that between \Nif mass and optical decline rate parameters noted in the literature \citep[e.g.][]{2007Sci...315..825M,2008ChJAA...8...71W,2014MNRAS.440.1498S}. Therefore adopting $t_2$ as a standardization parameter can lead to a smaller residual dispersion because $t_2$ is more strongly correlated to the physical driver of the luminosity, thus explaining our findings.

Future SN surveys are designed to provide multi-band data for a large sample of SNIa. Space-based observatories like Euclid and WFIRST will be equipped with NIR filters, which will allow us to observe SNIa in the $IYJH$ bands out to high-$z$. With such a configuration, we can expect measurements of $t_2$ for SNIa at $z$ \textgreater 0.5, with the view of using this quantity as an alternative standardization parameter to the optical decline rate. \citet{2014A&A...572A..80A} proposed a SN survey with LSST and the Euclid satellite out to z $\sim$ 1.5. With their survey parameters, they expect a total of $\sim$ 1700 SNe in the redshift range 0.75 \textless z \textless 1.5, a sizeable sample to test our standardization procedure at high-$z$. This would lead to  a better understanding of the physical parameters underlying the standardization procedure (e.g., $^{56}$Ni mass), to tests of the validity of the empirical stretch correction and to a reduction of the systematic error budget in SNIa cosmology. 

{\it Acknowledgements:} The authors would like to thank Kaisey Mandel for useful comments on an early draft. This work was supported by Grant ST/N000838/1 from the Science and Technology Facilities Council (UK). RT was partially supported by an EPSRC ``Pathways to Impact'' grant. DvD was supported by a Wolfson Research Merit Award (WM110023) provided by the British Royal Society and by Marie-Curie Career Integration (FP7-PEOPLE-2012-CIG-321865) grant provided by the European Commission. RT, DvD and HS were supported by a Marie-Skodowska-Curie RISE (H2020-MSCA-RISE-2015-691164) Grant provided by the European Commission.



\end{document}